\documentclass[amsfonts,amsmath,prd,preprint,nofootinbib]{revtex4}
\usepackage{epsfig,bbm,cancel,ulem}
\usepackage[breaklinks=true]{hyperref}
\usepackage[utf8]{inputenc}
\usepackage[english]{babel}
\usepackage{xcolor}
\usepackage{appendix}
\usepackage[papersize={8.5in,11in}]{geometry}

\begin{document}

\title{Thermodynamics of EiBI-AdS black holes with global monopole}

\author{A. M. Kusuma}
\author{B. N. Jayawiguna}
\author{H. S. Ramadhan}
\affiliation{Departemen Fisika, FMIPA, Universitas Indonesia, Depok, 16424, Indonesia. }
\def\changenote#1{\footnote{\bf #1}}

\begin{abstract}
It is well-known that black hole can be endowed with topological charge coming as a result of phase transition(s) in the early universe. It has also recently been revealed that such object might exist in one of the modified theory of gravity called the Eddington-inspired-Born-Infeld (EiBI) theory. Here we shall investigate the the possibility of phase transitions, and in general thermodynamic phenomena, of EiBI-anti-de Sitter (AdS) black hole with (topologically-charged) global monopole. This is the first work that applies the full Euclidean action formalism in this model. We provide a counterterm to cancel inifinities and argue that it is the most suitable among other possibilities. Our investigation reveals that the state variables obtained are found to obey the first law of the black hole mechanics and the Smarr's law for black holes with $\Lambda\neq0$. Related to the second feature, we obtained the forbidden range of parameter space for EiBI AdS black hole with global monopole, which corresponds to $\frac{-(1 - \Delta)}{\Lambda } \leq \kappa \leq -\frac{1}{\Lambda }$. The dependencies on $\Lambda$, the EiBI constant $\kappa$ and the global monopole charge $\eta$ of the state variables and state functions obtained manifests in a Schwarzchild AdS-like phase transition for black holes with parameters below the lower bound of the forbidden range, and could also manifest in a Schwarzschild-flat like phase behavior for black holes with parameters above the higher bound of the forbidden range.
\end{abstract}

\maketitle
\thispagestyle{empty}
\setcounter{page}{1}

\section{Introduction}
A series of phase transitions in the early universe might have produced global monopole(s) when global $SO(3)$ symmetry is spontaneously broken~\cite{Vilenkin}. These objects, along with other types of topological defects, are of great interest in cosmology as their existence might reveal information about the early history of our cosmos and the unification of fundamental interactions. The gravitational field of this object was first studied by Barriola and Vilenkin in 1989~\cite{Barriola:1989hx} that shows that, although the monopole exerts no gravitational force, nonetheless the metric is not asymptotically flat but suffers a deficit solid angle. When the monopole core is much smaller than the Schwarszchild radius, the solution describes a black hole eating up a global monopole~\cite{Dadhich:1997mh}. Ever since then, numerous studies have been devoted to the study of classical properties as well as cosmological signatures of global monopole black holes in various modifications. Black holes with non-canonical global monopole within the framework of General Relativity (GR) were studied, for example,  in~\cite{Jin:2007fz, Liu:2009eh, Prasetyo:2017rij}. Extensive works have also been done to investigating canonical black holes within the modified gravity models, for example in~\cite{Carames:2011uu} and the references therein. Recently, gravitational field of global monopole was studied in the framework of Eddington-inspired-Born-Infeld (EiBI) gravity~\cite{Lambaga:2018yzv}, a modified gravity model that revives an old proposal by Eddington and was proposed to solve singularity problems within GR~\cite{Banados:2010ix, Delsate:2012ky}.

Semiclassically, black holes radiate~\cite{Bekenstein:1973ur, Bardeen:1973gs} and, as a consequence, may undergo phase transition~\cite{Roychowdhury:2014cva}. The study of black hole phase transition began when Hut and, independently, Davies pointed out such possibility~\cite{HutP1977,Davies:1978zz}. Interests in this field of study then has been steadily increasing ever since Hawking and Page's seminal finding in 1983 that in AdS space, Schwarzschild black holes display phase behavior and thus undergo phase transitions \cite{Hawking:1982dh}. Adopting the Euclidean formulation, they found in Schwarzschild-AdS there are two kinds of black holes that can exist, the thermodynamically-stable large and the thermodynamically-unstable small black holes. By treating the AdS space as the background metric, it has also been found that at certain critical temperature there would be phase transition between the thermal AdS space and the stable large Schwarzschild AdS black hole state. From then on numerous mechanisms to explain the thermodynamics and critical behaviour of AdS black holes have been proposed. One of the most interesting finding between all the proposed discourses is the fact that the cosmological constant in black holes could be thought of as a varying parameter \cite{Kastor:2009wy}. Once the cosmological constant variation is taken into account, there shall be a modification in the first law of black hole mechanics so that its form would be consistent with the Smarr's relation. The cosmological constant variation itself then can be done in its associated form, namely as the variation of spacetime's negative pressure $P$. It has been established that for $4d$ black holes, the cosmological constant in terms of the natural units is defined as
\begin{equation}
 \Lambda = - 8 \pi P  = - \frac{3}{l^2}. \label{eq:pressure}
\end{equation}
Note that the $l$ above is the AdS radius constant. The existence of pressure indicates the presence of its conjugate form, thermodynamic volume $V$ \cite{Sekiwa:2006qj,Dolan:2010ha}. The introduction of $P$ and $V$ in the first law of black hole mechanics leads to a more comprehensive study of thermodynamic quantities of black holes, one of its most enterprising branch of study is now widely dubbed as black hole chemistry \cite{Kubiznak:2012wp,Kubiznak:2016qmn}.

In contrast to the extensive studies on thermodynamics and phase transitions of vacuum or electrically-charged black holes, the study of them charged with global monopole is relatively rare, and even much less for the case of modified gravity. The thermodynamics of global monopole black holes in $4d$ were investigated in~\cite{Jing:1993np,Yu:1994fy,Jensen:1995fz,Chen:2001jn,Deng:2018wrd,Kumara:2019xgt,Soroushfar:2020wch}, while their higher-dimensional generalization was discussed, for example, in~\cite{Ramadhan:2018gbf}. Semiclassical analysis of black holes in EiBI gravity were studied in~\cite{He:2016yuc, Ozen:2017uoe, Jayawiguna:2018tba}. It is therefore interesting to investigate the effect of global monopole to the critical behavior of AdS black holes with global monopole in EiBI gravity. That is the purpose of this work. To achieve that, this paper proceeds as follows. In the next section we review the static black hole with global monopole in EiBI gravity. In Sec. \ref{sec3} we present the Euclidean action formalism of EiBI AdS black hole with global monopole from which the corresponding partition function can be obtained. In Sec. \ref{sec4} we calculate the black hole's state variables and in Sec.~\ref{sec5} we present the local and global stability of this model by investigating its phase structure. Sec. \ref{sec6} is devoted to the conclusion and discussion. Note that throughout this work we have set $\hbar=c=G=1$, unless otherwise stated.

\section{EiBI-AdS global monopole (EiBI-GM-AdS)}
\label{sec2}

It is instructive to briefly review the EiBI global monopole solutions laid out in~\cite{Lambaga:2018yzv}. The action is
\begin{equation}
\label{action}
S = \frac{1}{8\pi \kappa} \int d^4 \tilde{x} \left( \sqrt{- \vert g_{\mu\nu}+ \kappa R_{\mu\nu}(\Gamma)\vert} -\lambda \sqrt{-g}\right) + S_{M}[g,\Phi_{M}].
\end{equation}
The matter Lagrangian is given by the global SO(3) Higgs field,
\begin{equation}
\mathcal{L} = \frac{1}{2}(\nabla\Phi^a)^2 -\frac{\sigma}{4}(\Phi^a \Phi^a - \eta^2)^2.
\end{equation} 
Employing the Palatini formalism, the gravitational field equations are
\begin{equation}
\label{2}
\sqrt{-q}q^{\mu\nu}=\lambda \sqrt{-g}g^{\mu\nu}-8\pi\kappa \sqrt{-g}T^{\mu\nu},
\end{equation}
and
\begin{equation}
\label{3}
q_{\mu\nu}= g_{\mu\nu}+\kappa R_{\mu\nu},
\end{equation}
for $g_{\mu\nu}$ and $q_{\mu\nu}$, respectively.

\begin{figure}[!htbp]
	\centering
	\begin{tabular}{cc}
		\includegraphics[height=6.5cm,keepaspectratio]{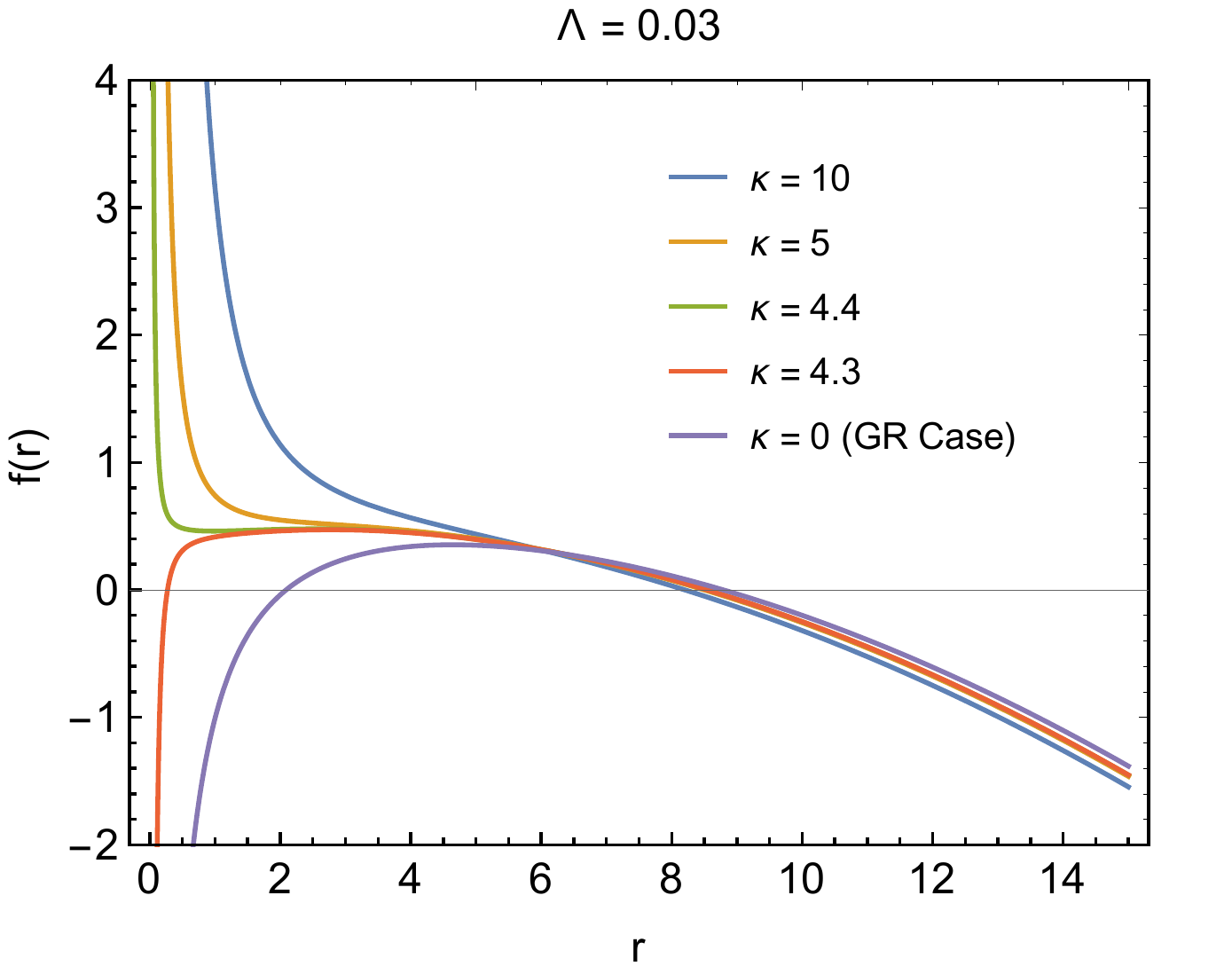} &		\includegraphics[height=6.5cm,keepaspectratio]{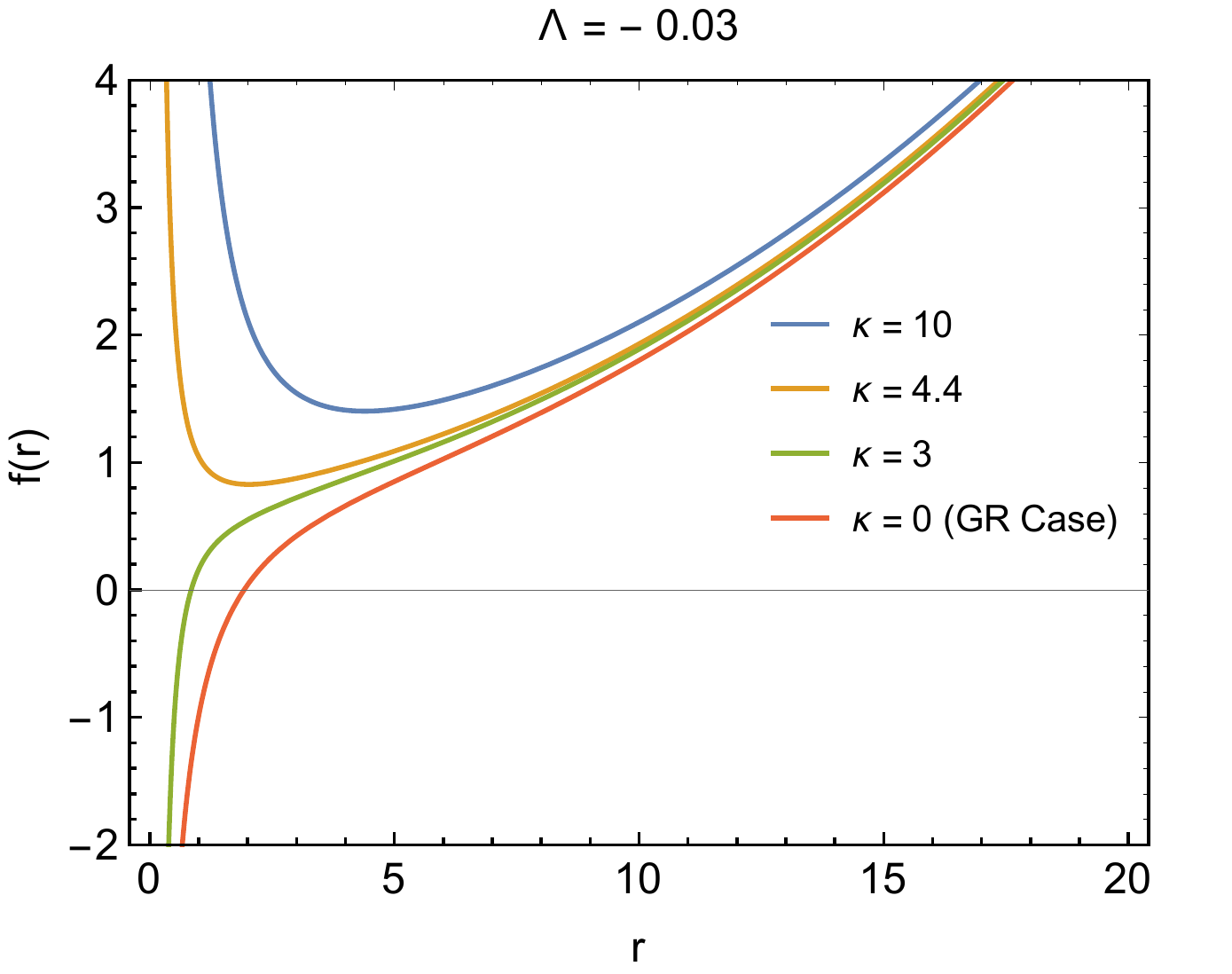}  
	\end{tabular}
	\caption{Profiles of $ f(r) $ for several values of $ \kappa $. In this plot, we set $ m=1 $.}
	\label{fig:frds}
\end{figure}

Assuming static, spherically symmetric configuration, we choose ansatz
\begin{eqnarray}
ds^2_{g}&=&-\psi^2(\tilde{r})f(\tilde{r})d\tilde{t}^2+\frac{1}{f(\tilde{r})} d\tilde{r}^2+\tilde{r}^2d\Omega^2_{2}, \label{5} \\
ds^2_{q}&=&-G^2(\tilde{r})F(\tilde{r})d\tilde{t}^2+\frac{1}{F(\tilde{r})} d\tilde{r}^2+H^2(\tilde{r})d\Omega^2_{2}, \label{6}
\end{eqnarray}
and 
\begin{equation}
\label{hedgehog}
\Phi^{a}= \phi(r)\eta\frac{x^a}{r},
\end{equation}
where $ds^2_g$ and $ds^2_q$ are the physical and auxiliary metrics, respectively. Outside the monopole core, the exterior energy momentum tensors are ($ \phi \approx1$) $ T^{\mu}_{\nu}=\eta^2/\tilde{r}^2~diag(-1,-1,0,0)$. Substituting it to Eq.~\eqref{2}, we get
\begin{eqnarray}
\label{HGF}
H^2 = \lambda \tilde{r}^2 + 8\pi \kappa \eta^2,~~ G= \lambda \psi,~~ F=\frac{f}{\lambda}.
\end{eqnarray} 
On the other hand, Eq.~\eqref{3} read
\begin{eqnarray}
\frac{2}{\kappa F} \left( \frac{\psi^2 f}{G^2 F} -1\right)&=& \frac{F''}{F}+\frac{2G''}{G}+\frac{3G'F'}{GF}+\frac{2F'H'}{FH} +\frac{4G'H'}{GH},\nonumber \\ \frac{2}{\kappa F}\left(\frac{F}{f}-1\right) &=& \frac{F''}{F}+\frac{2G''}{G}+\frac{4H''}{H}+\frac{2F'H'}{FH} +\frac{3F'G'}{FG},\nonumber \\ \frac{1}{\kappa F} \left(\frac{r^2}{H^2}-1\right) &=& -\frac{1}{H^2 F}+\frac{F'H'}{FH}+\frac{H'^2}{H^2} + \frac{H''}{H} + \frac{H'G'}{HG}.\nonumber\\
\label{4}
\end{eqnarray}
Solving Eqs.~\eqref{4} simultaneously, the solutions are
\begin{equation}
\psi(\tilde{r}) = \frac{\sqrt{\lambda} \tilde{r}}{\sqrt{\lambda \tilde{r}^2 + 8\pi\kappa \eta^2}},
\end{equation}
and
\begin{figure}[htbp]
	\centering\leavevmode
	\epsfysize=8cm \epsfbox{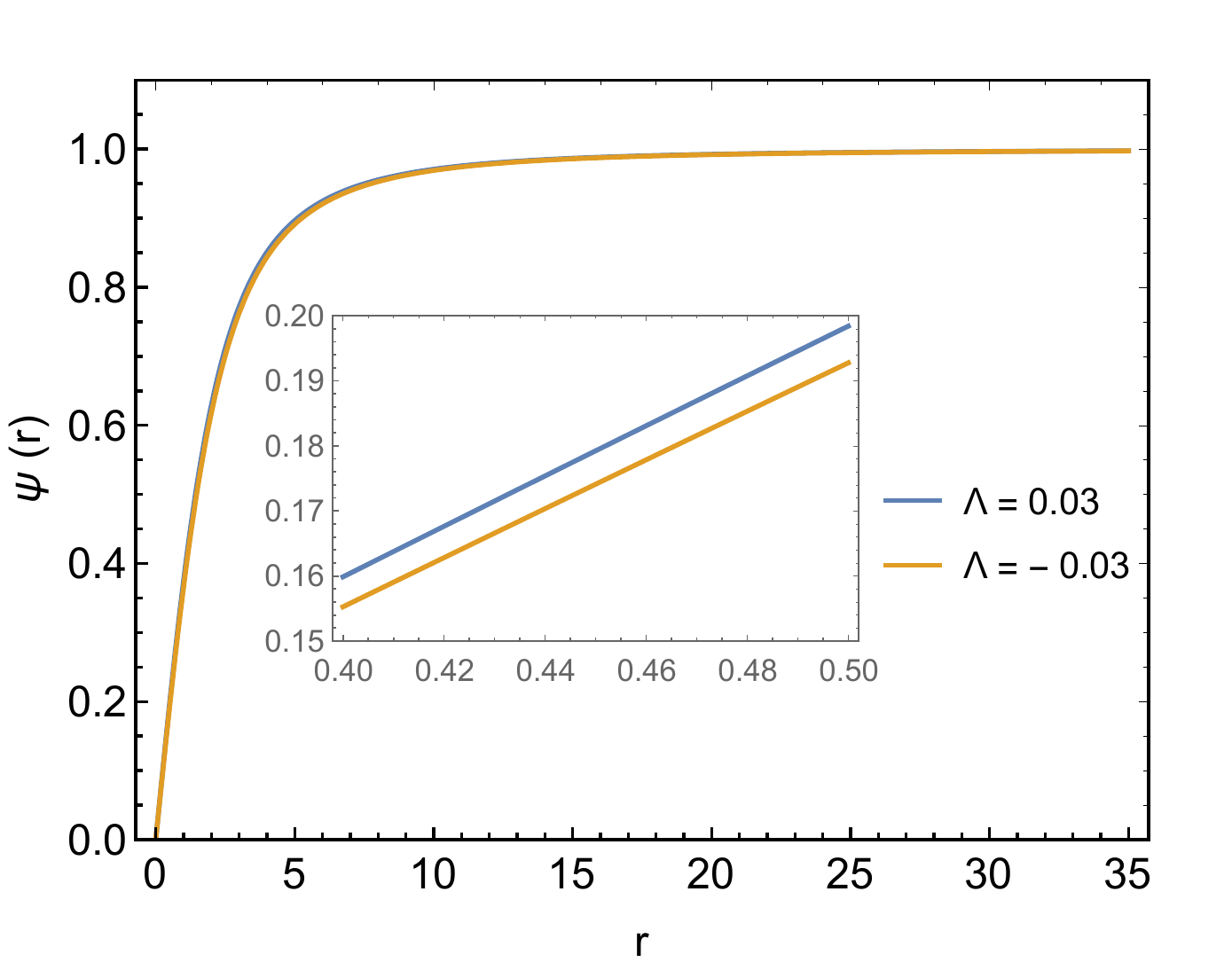}
	\caption {Plot of $ \psi(r) $ for $ \kappa=1 $ and $ \eta=0.5 $.}
	\label{fig:psi}
\end{figure}
\begin{equation}
f(\tilde{r}) = 1-\frac{2\tilde{m}\sqrt{\lambda}\sqrt{\lambda \tilde{r}^2 + \Delta \kappa}}{\lambda \tilde{r}^2} -\frac{\Lambda}{3} \tilde{r}^2-\left[\frac{2\Lambda}{3}+ \frac{(\tilde{r}^2-1)}{\tilde{r}^2}\right]\frac{\Delta}{\lambda}-\left(1+\frac{\Lambda}{3}\right) \frac{\Delta^2}{\lambda^2 \tilde{r}^2},
\end{equation}
with $ \Delta\equiv8\pi\eta^2 $. Defining $\tilde{\Delta}=\Delta/\lambda$ and applying the following rescaling prescriptions,
\begin{align}
    &\tilde{t} \rightarrow \frac{t}{\sqrt{1 - \tilde{\Delta}}},  &\tilde{r} \rightarrow r\sqrt{1 - \tilde{\Delta}}, &&\tilde{\Delta}\kappa \rightarrow (1 - \tilde{\Delta})\kappa, &&\tilde{m} \rightarrow m (1 - \tilde{\Delta})^{3/2}, \label{eq:rescale}
\end{align}
the line elements from Eqs. \eqref{5}-\eqref{6} are thus obtained to be
\begin{eqnarray}
ds^2_{g} &=& -\left[1- \frac{2m}{\sqrt{ r^2 + \kappa}} - \frac{1}{3}\Lambda(\kappa + r^2)  \right] dt^2 + \frac{ r^2}{ r^2 +\kappa} \left[1- \frac{2m}{\sqrt{ r^2 + \kappa}} - \frac{1}{3}\Lambda(\kappa + r^2)  \right]^{-1} dr^2 \nonumber \\ 
&& + r^2(1-\tilde{\Delta})d\Omega^2_{2}. \label{eq:rescaledfr} \\
ds^2_{q}&=&-\lambda\left[1 -\frac{2 m}{\sqrt{\kappa +r^2}}-\frac{1}{3} \Lambda  \left(\kappa +r^2\right)\right]dt^2+\frac{\lambda r^2}{\kappa +r^2} \left[1 -\frac{2 m}{\sqrt{\kappa +r^2}}-\frac{1}{3} \Lambda  \left(\kappa +r^2\right)\right]^{-1} dr^2 \nonumber \\
&& + \lambda (r^2 + \kappa) (1 - \tilde{\Delta})d\Omega^2_{2}. \label{eq:rescaledaux}
\end{eqnarray}
The profiles of $\psi(r)$ and $f(r)$ are shown in In Figs.~\ref{fig:psi}-\ref{fig:frds}.

\section{Euclidean action formulation} \label{sec3}

The complexified action integrals may serve as a starting point to obtain the state variables of black holes, as what has been first established by Gibbons and Hawking in 1977 \cite{Gibbons:1976ue}. For convenience, we shall relabel some components in Eqs. \eqref{eq:rescaledfr}-\eqref{eq:rescaledaux} as follows
\begin{eqnarray}
A(r) &\equiv& 1 -\frac{2 m}{\sqrt{r^2 + \kappa}}-\frac{1}{3} \Lambda  \left(r^2 + \kappa \right), \nonumber \\
B^{2}(r) &\equiv& \frac{r^2}{(r^2 + \kappa)}, \nonumber \\
C^2(r) &\equiv& (r^2 + \kappa).
\end{eqnarray}
Applying Wick rotation $(t \rightarrow i \tau)$ to the previously defined line elements, we obtain
\begin{eqnarray}
ds^2_{g} &=& A(r) d\tau^2 + B^2(r) A^{-1}(r) dr^2 + r^2(1-\tilde{\Delta})d\Omega^2_{2}, \label{eq:irescaledfr} \\
ds^2_{q}&=& \lambda A(r)d\tau^2 + \lambda B^2(r) A^{-1}(r) dr^2 + \lambda C^2(r) (1 - \tilde{\Delta})d\Omega^2_{2}. \label{eq:irescaledaux}
\end{eqnarray}
To construct the thermal states, we utilize the relation between the geometrical period of spacetime with the inverse temperature in Euclidean time \cite{Hawking:1982dh}. In our case, the Hawking temperature reads \cite{Visser:1992qh}
\begin{eqnarray}
T_{H}&=& \frac{\kappa}{2\pi} = \frac{1}{4\pi} \left[ \frac{\partial_r (g_{00})}{\sqrt{g_{00}g_{11}}} \right] \nonumber \\
&=& \frac{1}{4\pi} \left[\frac{1}{\sqrt{r^2_{+} + \kappa}} - \Lambda \sqrt{r^2_{+} + \kappa} \right]. \label{temp}
\end{eqnarray}

We should point out that in our case, the Hawking temperature obtained from the physical and auxiliary metric are exactly alike with each other. Thus, based on the result above the period of the black hole is given by
\begin{equation}
    \beta = 4\pi \left[ \frac{\sqrt{r^2_{+} + \kappa}}{1 - \Lambda(r^2_{+} + \kappa)} \right].
\end{equation}

The total Euclidean action considered would be of the following form
\begin{equation}
    I = I_b + I_{GHY} + I_{ct},
\end{equation}
where $I_b$ is the bulk action, $I_{GHY}$ is the Gibbons-Hawking-York boundary term and $I_{ct}$ is the counterterm.
The bulk action is the Wick rotated form of Eq. \eqref{action}, while $I_{GHY}$ and $I_{ct}$ serves to regulate the divergences that arise in the bulk action evaluation.
First we evaluate the bulk action with the line elements given in Eqs. \eqref{eq:irescaledfr}-\eqref{eq:irescaledaux}, which gives
\begin{eqnarray}
    I_b &=& \frac{1}{8\pi \kappa} \int d^4 x \left( - \sqrt{q} + \lambda \sqrt{g} - 8\pi\kappa \mathcal{L}_m \sqrt{g} \right) \nonumber \\
    &=& \frac{\beta\lambda}{2} \left[ \frac{\Lambda}{3} (r_+^2 + \kappa)^{3/2} - \frac{\Lambda}{3} (r_b^2 + \kappa)^{3/2} \right] (1 - \tilde{\Delta}) \label{eq:ibf}
\end{eqnarray}
Note that the result above can also be obtained by working with the unrescaled line elements given in Eqs. \eqref{5}-\eqref{6} on the bulk action evaluation, as long as we conduct appropriate rescaling prescription previously outlined in Eq. \eqref{eq:rescale} before the $t$ and $r$ elements integration.

Before we construct the $I_{GHY}$ for EiBI gravity, first let us note that the action given in Eq. \ref{action} could also be written in its alternative form \cite{Delsate:2012ky},
\begin{eqnarray}
    I_b &=& \frac{1}{16\pi\kappa} \int d^4 x \sqrt{-q} \left[ \kappa R - 2 + \left( q^{\mu\nu}g_{\mu\nu} - 2 \lambda \sqrt{\frac{(-g)}{(-q)}} \right) \right] + S_m, \nonumber \\
    &=& \frac{1}{16\pi} \int d^4 x \sqrt{-q} \left( R - \frac{2\Lambda}{\lambda} \right). \label{eq:altact}
\end{eqnarray}
We can see that expression has similar structure to the usual Einstein-Hilbert action. We then argue that the appropriate boundary term is the usual Gibbons-Hawking-York (GHY) form \cite{Gibbons:1976ue,York:1986it,Arroja:2016ffm}. Using the constructed trace of the extrinsic curvature from Eq. \eqref{eq:irescaledaux} it yields
\begin{eqnarray}
I_{GHY} &=& \frac{1}{8\pi}\int d^3 x \sqrt{h} K, \nonumber \\
&=& \frac{\beta \lambda}{2} \left( 3m + \Lambda(r_b^2 + \kappa)^{3/2} - 2 \sqrt{r_b^2 + \kappa} \right) (1 - \tilde{\Delta}). \label{eq:ighyf}
\end{eqnarray}

Besides the GHY term, we also need to subtract the contribution of the black hole's background space. It might be tempting to immediately construct the counterterm in such a way that it comes from 3D form of Eq \eqref{eq:altact}, but such approach would not yield an effective counterterm, as we show in the Appendix~\ref{appx}. Instead, we shall take the advantage of how the alternative form of the EiBI GM AdS action is similar to the usual Einstein AdS action. Consider the second term of \eqref{eq:altact}. Using the rescaled line element given in Eq. \eqref{eq:irescaledaux}, we can see that the AdS background in EiBI gravity could be written in the following form
\begin{eqnarray}
    I_{EiBI AdS} &=& \frac{1}{16\pi\lambda} \int d^4 x \sqrt{\lambda^4 A(r) B^2(r) A^{-1}(r) C^{4}(r) (1 - \tilde{\Delta})^2 \sin^2 \theta} \left( 2 \Lambda \right), \nonumber \\
    &=& \frac{\lambda}{16\pi} (1 - \tilde{\Delta}) \int d^4x \sqrt{g} \left( 2\Lambda \right). \label{eq:ieibiads}
\end{eqnarray}
This looks similar, up to some overall constant, to the usual AdS term found in Einstein-Hilbert (EH) action,
\begin{equation}
    I_{EH AdS} = \frac{1}{16\pi} \int d^4x \sqrt{g} \left( 2\Lambda \right).
\end{equation}

Due to the similarity of the form above with the usual EH action, it is justifiable to use the same counterterm prescription as used in AdS black holes with GR framework. In the search for the counterterm prescription, we found that the most appropriate and effective approach to compute the counterterm action is by utilizing the generalized form of the Balasubramanian-Kraus counterterm (BK) action~\cite{Emparan:1999pm,Balasubramanian:1999re}. For $4D$ AdS black holes the counterterm is given as follows
\begin{equation}
    I_{ct} = \frac{1}{8\pi} \int d^3 x \sqrt{h^o} \left[ \frac{2}{l} + \frac{l}{2} R - \frac{l^3}{2} \left( R_{\mu\nu} R^{\mu\nu} - \frac{3}{8}R^2 \right)  \right]. \label{eq:ict}
\end{equation}

Thus, we will use the counterterm prescription for AdS black holes given in Eq. \eqref{eq:ict} by treating the $\lambda(1 - \tilde{\Delta})$ as a constant. The Ricci scalar and tensor shall be constructed from the following boundary metric
\begin{equation}
    h^o_{ij} = \left\{ A_o(r_b), C^{2}(r_b), C^{2}(r_b) \sin^2 \theta \right\},
\end{equation}
with $A_o$ is the modified AdS metric in EiBI gravity and we will define it as follows
\begin{equation}
    A_o (r) = 1 - \frac{\Lambda}{3}(r^2 + \kappa).
\end{equation}
Using the quantities above, the counterterm gives
\begin{eqnarray}
I_{ct} &=& \frac{\lambda}{8\pi} (1 - \tilde{\Delta}) \int d^3 x_o \sqrt{h_o} \left[ \frac{2}{l} + \frac{l}{2} R - \frac{l^3}{2} \left( R_{\mu\nu} R^{\mu\nu} - \frac{3}{8}R^2 \right)  \right], \nonumber \\
&=& \frac{\lambda \beta_o \sqrt{A_o(r_b)}}{2} \left\{ -\frac{l^3}{(r_b^2 +\kappa)}+\frac{3 l^3}{4 (r_b^2 +\kappa)}+\frac{2 (r_b^2 +\kappa)}{l}+l \right\} (1 - \tilde{\Delta}). \label{eq:ict1}
\end{eqnarray}


Rescale the period of the AdS background so that it matches with that of the black hole,
\begin{equation}
    \beta_o \sqrt{A_o(r_b)} = \beta \sqrt{A(r_b)},
\end{equation}
and with some approximation we shall obtain
\begin{equation}
    \beta_o = \beta \left[-\frac{l^3 \left(\sqrt{r_b^2 +\kappa}-2 m\right)^2}{8 \left(r_b^2 +\kappa\right)^{5/2}}+\frac{l \left(\sqrt{r_b^2 +\kappa}-2 m\right)}{2 \left(r_b^2 +\kappa\right)}+\frac{\sqrt{r_b^2 +\kappa}}{l} \right].
\end{equation}

Substituting the results above into Eq. \eqref{eq:ict1} and omitting the null terms (terms that becomes zero as $r_b\rightarrow\infty$) as we revert the $l$ constant back into $\Lambda$, leave us with the following results
\begin{equation}
    I_{ct} = \frac{\beta \lambda}{2} \left[ -2m + 2 \sqrt{r_b^2 + \kappa} - \frac{2\Lambda}{3} \left( r_b^2 + \kappa \right)^{3/2} \right] (1 - \tilde{\Delta}). \label{eq:ictf}
\end{equation}
Combining the results that we have obtained so far from Eqs. \eqref{eq:ibf}, \eqref{eq:ighyf} and \eqref{eq:ictf} yields the total Euclidean action as follows
\begin{equation}
    I = \frac{\beta\lambda}{2} \left[ \frac{\Lambda}{3} (r_+^2 + \kappa)^{3/2} + m \right] (1 - \tilde{\Delta}).
\end{equation}
The value of $m$ can be obtained from $f(r_+)=0$ and gives
\begin{equation}
    m = -\frac{1}{6} \sqrt{r_+^2 + \kappa} \left[ \Lambda(r_+^2 + \kappa) - 3 \right]. \label{eq:m}
\end{equation}
Thus, the total Euclidean action can be written again as follows
\begin{equation}
    I = \frac{\beta\lambda}{2} \left[ \frac{\Lambda}{6} (r_+^2 + \kappa)^{3/2} + \frac{1}{2} \sqrt{r_+^2 + \kappa} \right] (1 - \tilde{\Delta}).
\end{equation}

\section{State Variables, First law  and Smarr's formula}\label{sec4}

Using the total Euclidean action, we could compute state variables and state functions of the black hole. First we shall evaluate the entropy of the system, which gives
\begin{eqnarray}
    S &=& \beta \left( \frac{\partial I}{\partial \beta} \right)_A - I, \nonumber \\
    &=& \beta \left( \frac{\partial I}{\partial r_+} \right)_A \Big/ \left( \frac{\partial \beta}{\partial r_+} \right)_A - I, \nonumber \\
    &=& \lambda \pi (r_+^2 + \kappa) (1 - \tilde{\Delta}). \label{s}
\end{eqnarray}
We can see that the entropy for our case is different from that of the ordinary Einstein-Hilbert gravity, due to the existence of the $\lambda$, $\kappa$ and $\Delta$. Nevertheless, as $\kappa\rightarrow0$ we can see that the result reduces to that of the Bekenstein-Hawking entropy of a black hole that is endowed with global monopole \cite{Deng:2018wrd}. Note that the form given in Eq. \eqref{s} can also be obtained from the known relation between the black hole's entropy and the event horizon area calculated from the auxiliary metric
\begin{equation}
    S = \frac{A}{4} = \frac{1}{4} \int \sqrt{q_{\theta\theta}q_{\phi\phi}} d\theta d\phi.
\end{equation}

The {\it Generalized Second Law} (GSL) of black hole mechanics states that the entropy of a black hole must always increase, $S \geq 0$ \cite{Bekenstein:1974ax,Wald:1999vt}. For the entropy obtained in Eq. \eqref{s}, it seems certain value of $\kappa$ would yield decreasing amount of entropy, which clearly violates GSL. Thus we proceed to find the allowed set of parameters for $\kappa$ by applying the GSL requirement to the entropy expression, which yields the following result
\begin{eqnarray}
    1 - \tilde{\Delta} \geq 0, \nonumber \\
    1 - \frac{\Delta}{1 + \kappa \Lambda} \geq 0.
\end{eqnarray}
Setting the conditions that $0 \leq \Delta < 1$ (based on the critical value for monopole charge \cite{Barriola:1989hx,Prasetyo:2015bga}) and $\Lambda < 0$, we obtain the following range that should be considered as the requirement that the hole is thermodynamically feasible
\begin{align}
&\kappa \leq \frac{-(1 - \Delta)}{\Lambda } & \text{or} && \kappa >-\frac{1}{\Lambda }. \label{eq:parameterspace}
\end{align}
We can see the first condition on the equation above may serve as some sort of the lower bound while the other one acts as the upper bound. The region between these two boundaries $\left(\frac{-(1 - \Delta)}{\Lambda } \leq \kappa \leq -\frac{1}{\Lambda } \right)$ will be considered as the forbidden region for our thermodynamic analysis.

The ADM mass from the euclidean action is computed as follows
\begin{eqnarray}
    M &=& \left( \frac{\partial I}{\partial \beta} \right), \nonumber \\
    &=& - \frac{\lambda (1 - \tilde{\Delta})}{6} \sqrt{r_+^2 + \kappa} \left[ \Lambda(r_+^2 + \kappa) - 3 \right]. \label{eq:adm1}
\end{eqnarray}
Comparing the result above with Eq. \eqref{eq:m}, we can see that the ADM mass can also be expressed in the following form
\begin{equation}
    M = \lambda (1 - \tilde{\Delta}) m. \label{eq:admmass}
\end{equation}
We can also see that once again there is an explicit dependence on $\lambda$, and consequently $\kappa$, on the ADM mass. As $\kappa\rightarrow0$ the result would still reduces to the known value in EH gravity \cite{Deng:2018wrd}.

In 2008, Kastor~\cite{Kastor:2008xb} constructed a new Komar integral for black holes with $\Lambda\neq0$ by defining an antisymmetric 2-form potential $\omega^{\mu\nu}$ such that
\begin{equation}
M = \int_{\partial \Sigma} dS_{\mu\nu} \left( \nabla^{\mu}\xi^{\nu} +  \Lambda \omega^{\mu\nu} \right),
\end{equation}
where $\omega^{\mu\nu}$ satisfies $\nabla_{\mu}\omega^{\mu\nu}=0$ and $ \xi^{\mu}\equiv\lbrace 1,0,0,0 \rbrace $ is the time-invariant Killing vector. The equation above can also be used to obtain the ADM mass given in Eq. \eqref{eq:admmass}.
In the following year Kastor {\it et al} found that there is a consistency between the derivation of Smarr's formula for AdS black holes from the newly defined Komar integral and the scaling argument (the method which was actually used by Smarr in his original proposal), if the variation of $\Lambda$ is also taken into account \cite{Kastor:2009wy}. The modified Smarr's formula could be expressed as follows
\begin{equation}
    M = 2TS - 2VP. \label{eq:smarr}
\end{equation}
The volume term $V$ above is defined by the relation between $P$ and $\Lambda$ given in Eq. \eqref{eq:pressure}, and is given by
\begin{equation}
    V = \left( \frac{\partial M}{\partial P} \right)_A.
\end{equation}
The first law of black hole mechanics as the variation of $\Lambda$ is taken into account can then be defined as follows
\begin{equation}
dM = T dS + V dP. \label{eq:firstlaw}
\end{equation}
Employing Eqs \eqref{eq:pressure} and \eqref{eq:admmass} to the equation above, the conjugate volume is obtained to be
\begin{equation}
    V=\frac{4\pi}{3}(1-\tilde{\Delta})(r^2_{+} + \kappa)^{3/2}. \label{eq:thermovol}
\end{equation}

It can be verified that the state variables that we have obtained so far for our model, namely Eqs. \eqref{temp}, \eqref{s}, \eqref{eq:admmass} and \eqref{eq:thermovol}, along with the definition of $P$ given in Eq. \eqref{eq:pressure} would obey \eqref{eq:smarr} and \eqref{eq:firstlaw}.

\section{The thermodynamic stability of an AdS EiBI black hole with global monopole}\label{sec5}

\begin{figure}[htbp]
	\centering\leavevmode
	\epsfysize=8.5cm \epsfbox{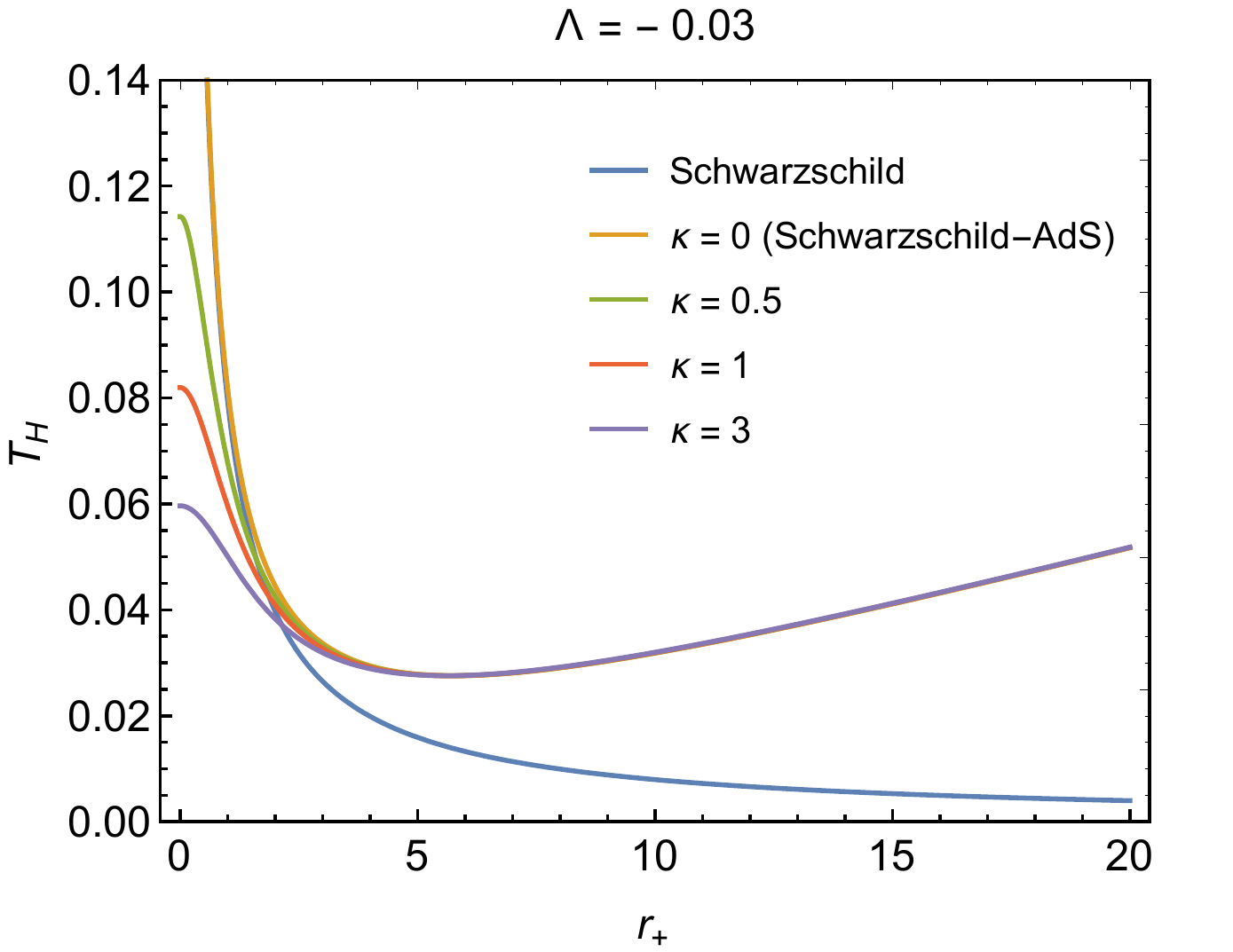}
	\caption {A typical plot of Hawking temperature $T_{H}$ versus radius $ r_{+} $.}
	\label{fig:adstemp}
\end{figure}

After obtaining the relevant state variables of the EiBI-GM-AdS black holes, the thermodynamical stability can be investigated. We start with the local stability, which can be done by analyzing the Hawking temperature and the specific heat of the configuration. The plot of Hawking temperature $T_{H} $ vs the event horizon $ r_{+} $ is shown in Fig.~\ref{fig:adstemp}. There is a discontinuous transition from lower ($ \kappa=0 $) to higher $ (\kappa>0) $ EiBI parameter. In the Schwarzschild case ($ \kappa=0 $ and $ \Lambda=0 $), the Hawking temperature depends only on its mass or event-horizon radius and slowly decreases as the radius gets larger. When $ \kappa = 0 $ (with nonvanisihing cosmological constant), the Hawking temperature becomes similar to the well known Schwarzschild-AdS. In this case, the black hole has minimum temperature at $ r_{0}=i / \sqrt{\Lambda} $. This condition forces $ T_{0}= \frac{-i\sqrt{\Lambda}}{2\pi} $, where the cosmological constant must have a negative value (AdS), $ \Lambda = -|\Lambda| $. From this plot, we can also infer that for $T<T_{0}$ the spacetime is filled with pure radiation and hence there is no AdS black hole. For $ T>T_{0}, $ AdS black holes tend to have two different conditions. When EiBI parameter $ \kappa $ gets larger, we still have a same behavior as Schwarzschild-AdS but with the nonzero Hawking temperature at the origin. We shall discuss in detail about this conditions and also the stability on specific heat.
\begin{figure}[htbp]
	\centering\leavevmode
	\epsfysize=6 cm \epsfbox{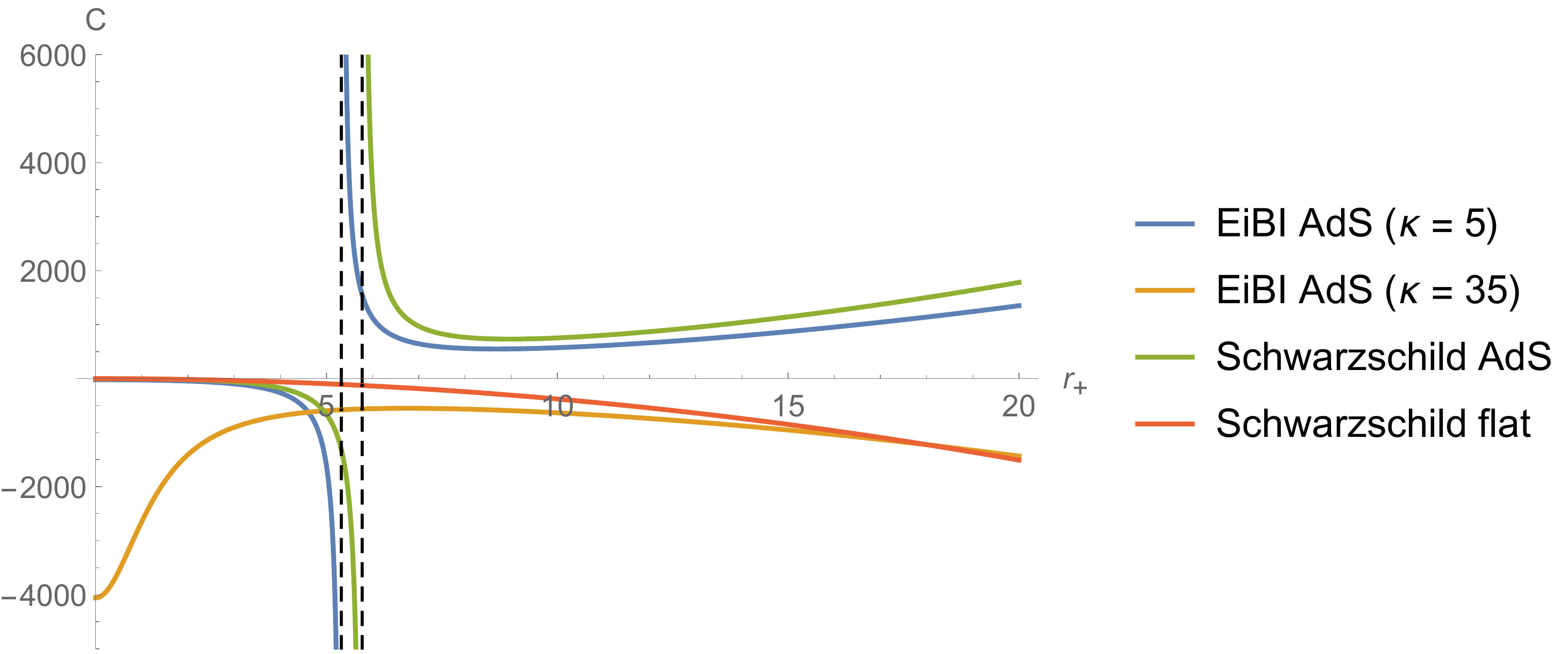}
	\caption{The heat capacity $ C $ vs $r_+$ for $\Lambda=-0.03$ and $ \Delta=0.4 $.}
	\label{fig:cqfig}
\end{figure}

The Hawking temperature in Eq.~\eqref{temp} indicates that there is a possible transition from small to large black hole. In detail, we can use the specific heat $C$ and free energy $F$ to investigate the stability and phase transitions. Combined with the entropy \eqref{s}, the specific heat $ C $ can be obtained:
\begin{eqnarray}
C= T_{H}~ \frac{\partial S}{\partial T_{H}}= - \frac{2\pi\lambda(r_+^2 + \kappa)(-1 + \Lambda r_+^2 + \kappa\Lambda)}{(1 + \Lambda r_+^2 + \kappa \Lambda)} (1 - \tilde{\Delta}). \label{eq:cq}
\end{eqnarray}
Fig.~\ref{fig:cqfig} shows the specific heat described in Eq. \eqref{eq:cq}, compared to those of Schwarzschild black holes in AdS and asymptotically flat spacetimes. Their stability depend on the value of $C$, where positive sign ($ C>0 $) indicates the black holes are stable and $ C<0 $ means unstable.

In general, the specific heat plot of an EiBI-GM-AdS black hole that follows the first parameter space given in Eq. \eqref{eq:parameterspace} would have diverging points separating the two distinct phases, illustrated with the blue-colored plot in Fig. \ref{fig:cqfig}. The form is completely similar to that of the Schwarzschild-AdS black hole (plotted as the green-colored plot in Fig. \ref{fig:cqfig}), which tells us that the nature of the phase transition taking place between these two configurations black holes is the same; there would be a phase transition taking place between the unstable small black hole (SBH) and the stable large black hole (LBH). Meanwhile, for EiBI GM AdS black holes that follow the second parameter space in Eq. \eqref{eq:parameterspace}, we observed the specific heat would always be completely negative, reminiscent with the usual flat Schwarzschild black hole configuration. The main difference between the two configurations is the existence of unstable equilibrium for the EiBI-GM-AdS, while for flat Schwarzschild black hole there would be none (due to the quadratic nature of the function) whereas for EiBI GM AdS configuration there is one visible.

With the results obtained so far, we have been able to analyze the local stability of the holes. Now we proceed to extend the analysis by determining the global stability of the configuration in order to study the phase of a system that corresponds to the global maximum of the system's total entropy, in other words phase that minimize the free energy used would be the preferred one \cite{Monteiro:2010cq}. The free energy in black hole mechanics is defined as follows,
\begin{equation}
    F = \frac{I}{\beta}.
\end{equation}
Using the previously obtained results for the thermodynamic variables, the free energy as a function of the black hole's event horizon is obtained to be
\begin{equation}
F   =  \frac{\lambda}{2} \left[ \frac{\Lambda}{6} (r_+^2 + \kappa)^{3/2} + \frac{1}{2} \sqrt{r_+^2 + \kappa} \right] (1 - \tilde{\Delta}).
\end{equation}
Note that the results above agree with $F = M - TS$, using expressions previously obtained in Eqs \eqref{temp}, \eqref{s} and \eqref{eq:adm1}.
\begin{figure}[htbp]
	\centering\leavevmode
	\epsfysize=6cm \epsfbox{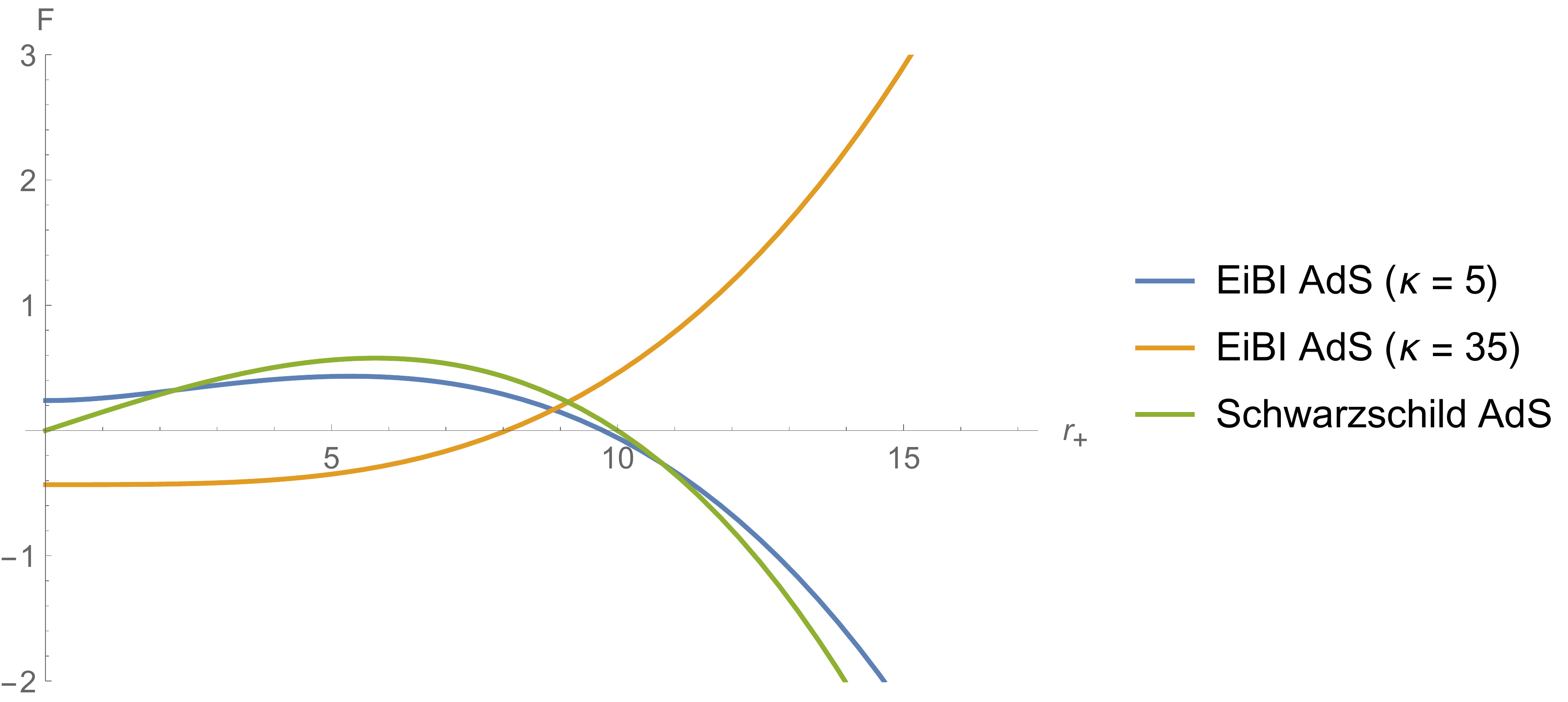}
	\caption {Figure that represents the on-shell free energy of AdS holes as a function of event horizon with different configurations. In this plot, we have set $\Lambda=-0.03$ and $\Delta=0.4 $.}
	\label{fig:helmonads}
\end{figure}

The plots the free energy as a function of event-horizon are shown in Fig. \ref{fig:helmonads}. Echoing the earlier results presented in the specific heat results, we observed similar features between the EiBI-GM-AdS with $\kappa$ that follows the first parameter space given in Eq. \eqref{eq:parameterspace} with the Schwarzschild AdS black holes. The main difference between these two configurations are the fact that EiBI-GM-AdS starts with non zero free energy. At certain point, increasing the amount of $\kappa$ eventually renders the free energy configuration to be completely positive-valued. This drastic change of the free energy profile starts after the lower bound of the "forbidden region" of the EiBI parameter obtained previously from the applied GSL requirement to the entropy expression of the holes. Above this value, $F$ would abruptly change its profile into an ever-increasing value of free energy with no critical points, which leads us to consider any configuration above this particular value would not be thermodynamically stable.

\section{Conclusion} \label{sec6}

To summarize, we consider the 4d (A)dS black hole with a global monopole~\cite{Lambaga:2018yzv} as a test case to apply Euclidean formalism and derive the corresponding thermodynamical properties in EiBi gravity. To the best of our knowledge this is the first approach to do so in this type of modified gravity model. In non-canonical GR one is faced with ambiguity in choosing the appropriate boundary and counter terms to render the Euclidean action finite. In this work we argue that the most suitable respective terms are the GHY and the generalized KS forms.

Once the finite Euclidean action is obtained, it is easy to verify that the state variables satisfy the black hole first law and the Smarr formula. We obtained the parameter space in terms of $\kappa$, $\eta$ and the cosmological constant $\Lambda$ that would allow a thermodynamically feasible configuration. The phase structure of EiBI AdS with global monopole is found to be quite similar to that of Schwarzchild-AdS black hole endowed with global monopole, as long as $\kappa \leq \frac{-(1 - \Delta)}{\Lambda }$. On the other hand, once $\kappa >-\frac{1}{\Lambda }$, a Schwarzschild-flat like phase behavior shall be observed in our configuration.

All in all, we have been able to examine the nature of the phase transition taking place in the EiBI AdS black hole endowed with global monopole. Naturally, further studies on the nature of the phase transition in the electrically-charged black hole endowed with global monopole in EiBI gravity would seem to be an interesting inquiry to pursue.

\acknowledgments

We thank Ihsan Fauzi for the help with the preliminary manuscript. We also thank the anonymous referee for the constructive suggestions. This work is funded by PUTI Q2 grant under the contract No.~NKB-1654/UN2.RST/HKP.05.00/2020 and Hibah Riset PPI Q1 No. NKB-583/UN2. RST/HKP.05.00/2021.

\section*{Data Availability Statement}

Data sharing is not applicable to this article as no data sets were generated or analyzed during the current study.


\appendix
\section{Alternative Form of the Counterterm Action}
\label{appx}
One of the pivotal steps of Euclidean action calculation is the counterterm action evaluation. There are many approaches that can be done to calculate counterterm actions \cite{Gibbons:1976ue,York:1986it,Balasubramanian:1999re,Mann:1999pc}, and in this work we obtain finite result by utilizing the relatively simple form of counterterm action for AdS black holes \cite{Emparan:1999pm}. This is not a unique choice due to the ambiguity of choosing it in non-canonical gravity. In this section we shall show several attempts in constructing the counterterm action, and why they fail\footnote{We thank the anonymous PRD referee for suggesting these models for us to consider.}. The first approach is motivated by the known results in AdS black holes thermodynamics with GR structure, in which $I_{GHY}$ would eventually give vanishing contribution due to the nature of AdS. Thus, to normalize the divergent term arising from the bulk action one could construct a counter term from the Euclideanized form of the pure AdS geometry. In this approach, the integrand is the physical metric tensor, not the induced metric. Looking back at Eq. \eqref{eq:altact}, since there is similarity between EiBI and EH action, it is tempting to construct the counterterm in such a way that the form becomes as follows
\begin{equation}
     I_{ct_1} = \frac{1}{16\pi\lambda} \int d^4 x \sqrt{q} ( 2\lambda).
\end{equation}
Using the known result in AdS thermodynamics that for even-dimensional black holes in GR framework $r_+=0$ \cite{Gibbons:2004ai}, we then obtain
\begin{eqnarray}
    I_{ct_1} &=& \frac{\omega\beta_0 \Lambda \lambda}{8\pi} \left\{ \int_{0}^{r_b} dr \left[ r \sqrt{r^2 + \kappa} \right] \right\}(1 - \tilde{\Delta}), \nonumber \\
    &=& \frac{\lambda \beta_0 \Lambda}{2} \left\{ \frac{1}{3} \left(r^2 + \kappa\right)^{3/2} \Big\vert_0^{r_b} \right\}(1 - \tilde{\Delta}). \label{eq:ictalt1a}
\end{eqnarray}
Rescaling the period to the metric results in the following form
\begin{equation}
    \beta_0 = \beta \left[1 + \frac{l^2 \left(\sqrt{r_b^2 + \kappa}-2 m\right)}{2 \left(r_b^2 + \kappa\right)^{3/2}}  \right]. \label{eq:beta0match}
\end{equation}
Plugging the result above into Eq. \eqref{eq:ictalt1a} gives
\begin{equation}
    I_{ct} = \frac{\beta\lambda}{2} \left\{m - \frac{(r_b^2 + \kappa)}{2} + \frac{\Lambda}{3}(r_b^2 + \kappa)^{5/2} - \frac{\kappa^{3/2}}{3} \right\}(1 - \tilde{\Delta}). \label{eq:ictalt1}
\end{equation}
The result above actually reduces the divergent term that arise in the bulk action, yet we can see there is an extra divergent term that comes up in Eq \eqref{eq:ictalt1}, $-\left(r_b^2+\kappa\right)/2$. Thus the regular AdS counterterm approach does not seem to work completely in our case.

Next, we attempt to slightly modify the first approach that we have taken. If the counterterm was to be calculated as the rescaled to the AdS space version of the bulk action, then we would have
\begin{eqnarray}
    I_{ct_2} &=& \frac{1}{8\pi \kappa} \int d^4 x \left( \sqrt{g_{\mu\nu}+ \kappa R_{\mu\nu}} -\lambda \sqrt{g_{\mu\nu}}\right), \nonumber \\
    &=& \frac{1}{8\pi \kappa} \int d^4 x \left( \sqrt{q_{\mu\nu}} -\lambda \sqrt{g_{\mu\nu}}\right).
\end{eqnarray}
Similar to the previous approach, by taking advantage of $r_+=0$ in even dimensional AdS counterterm calculation, we obtain
\begin{eqnarray}
    I_{ct_2} &=& \frac{\omega \beta_0 \lambda}{8\pi \kappa} \int_0^{r^b} dr \left[ \lambda B(r) C^2(r) - B(r) r^2 \right](1 - \tilde{\Delta}), \nonumber \\
    &=& \frac{\beta_0 \lambda}{2\kappa} \left\{ \kappa \left[ \sqrt{r^2 + \kappa} + \frac{\Lambda}{3}(r^2 + \kappa)^{3/2} \right] \Bigg\vert_0^{r_b} \right\}(1 - \tilde{\Delta}), \nonumber \\
    &=& \frac{\beta_0 \lambda}{2} \left[ \sqrt{r_b^2 + \kappa} + \frac{\Lambda}{3}(r_b^2 + \kappa)^{3/2} - \sqrt{\kappa} - \frac{\Lambda}{3}\kappa^{3/2} \right] (1 - \tilde{\Delta}).
\end{eqnarray}
Using the $\beta_0$ rescaling obtained in Eq. \eqref{eq:beta0match} and substituting it into the form above would yield the following form (after the null terms are eliminated $r_b \rightarrow \infty$)
\begin{eqnarray}
    I_{ct_2} &=& \frac{\beta \lambda}{2} \left[ m - \frac{\sqrt{\kappa}}{3}(\lambda +2) + \frac{\sqrt{r_b^2 + \kappa}}{2}  + \frac{\Lambda}{3}(r_b^2 + \kappa)^{3/2}\right] (1 - \tilde{\Delta}).
\end{eqnarray}
Once again we see that the regular AdS counterterm approach does not completely normalizes the Euclidean action, instead we are left with more divergent terms. This is one of the reason why we choose to use the generalized BK boundary counterterm action, since they wonderfully work well in eliminating the divergence.

Our last attempt is by considering another approach, namely by reinterpreting the 3D Dirac-Born-Infeld action as the counterterm~\cite{Ozen:2017uoe,Gullu:2010pc}. This is motivated by the similarity of EiBI gravity with the new massive gravity \cite{BeltranJimenez:2017doy}. The 3D Dirac-Born-Infeld action itself is given by
\begin{equation}
    I_{DBI} = - \frac{m^2}{4\pi G_3} \int_{\mathcal{M}} d^3 x \left[ \sqrt{-\Big\vert g_{\mu\nu}+ \frac{\sigma}{m^2}\mathcal{G}_{\mu\nu} \Big\vert} - \left(1 - \frac{\lambda_0}{2}\right)\sqrt{-\vert g_{\mu\nu} \vert} \right]. \label{eq:dbiact}
\end{equation}
In the form above $m$ is a spin-related mass paramater. $G_3$ is the gravitational constant and $\sigma$ is term related to the helicity of Einstein-Hilbert term \cite{Bergshoeff:2009aq}. $\mathcal{G}_{\mu\nu}$ is the usual Einstein tensor $(\mathcal{G}_{\mu\nu}\equiv R_{\mu\nu} - \frac{1}{2}g_{\mu\nu}R)$. In Dirac-Born-Infeld gravity, the effective cosmological constant is expressed as
\begin{equation}
    \Lambda = \sigma m^2 \lambda_0 \left(1 - \frac{\lambda_0}{2}\right).
\end{equation}
Our goal is to reinterpret Eq.~\eqref{eq:dbiact} with appropriate terms in EiBI gravity, so that we may evaluate it as the counterterm action for our model. First we can see that the $m^2$ term in \eqref{eq:dbiact} is dimensionally similar to $\kappa$ in EiBI gravity, in which both of them are inversely proportional to the cosmological constant. Based on this observation if we were to reinterpret the form given in Eq \eqref{eq:dbiact}, it is tempting to translate $m^2 \rightarrow \kappa$. Next we may set $\sigma = -1$ since we are using the regular form of EH action. $G_3$ can be set to 1 so that eventually we obtain the following form
\begin{eqnarray}
    I_{ct_3} &=& \frac{1}{8\pi\kappa} \int d^3 x \left[ \sqrt{\vert g_{\mu\nu} + \kappa \mathcal{G}_{\mu\nu} \vert} + \lambda\sqrt{\vert g_{\mu\nu}\vert} \right], \nonumber \\
    &=& \frac{1}{8\pi\kappa} \int d^3 x \left[ \sqrt{\Big\vert g_{\mu\nu} + \kappa \left( R_{\mu\nu} - \frac{1}{2}g_{\mu\nu}R \right) \Big\vert} + \lambda\sqrt{\vert g_{\mu\nu}\vert} \right], \nonumber \\
    &=& \frac{1}{8\pi\kappa} \int d^3 x \left[ \sqrt{\Big\vert q_{\mu\nu} - \frac{\kappa}{2}g_{\mu\nu}R \Big\vert} + \lambda\sqrt{\vert g_{\mu\nu}\vert} \right]. \label{eq:ictalt3a}
\end{eqnarray}
The Ricci scalar from 3-D $q_{\mu\nu}$ could be constructed as
\begin{equation}
    R = \frac{2}{(1-\tilde{\Delta}) \lambda  \left(\kappa +r^2\right)}.
\end{equation}
Note that the form above is different from the one used in Eqs \eqref{eq:ict} and \eqref{eq:ict1}, since in this current approach we could not take advantage of the similarity between EiBI and EH action ($\lambda$ and $1 - \tilde{\Delta}$ could not be treated as mere constants in the counterterm evaluation anymore).
Evaluating Eq \eqref{eq:ictalt3a} gives
\begin{eqnarray}
    I_{ct} &=& \frac{\omega\beta_0}{8\pi\kappa} \left\{\sqrt{\frac{\left(6 m+\sqrt{\kappa +r_b^2} \left(\kappa  \Lambda +\Lambda  r_b^2-3\right)\right) \left(\kappa  \left(1 - (1 - \tilde{\Delta}) \lambda ^2\right)-(1-\tilde{\Delta}) \lambda ^2 r_b^2\right)}{3 (1 - \tilde{\Delta}) \lambda  \left(\kappa +r_b^2\right)^{3/2}}} \right. \nonumber \\
    && \left. \times \left[ \kappa  \lambda +r_b^2 \left(\lambda - \frac{\kappa }{(1-\tilde{\Delta}) \lambda  \left(\kappa +r_b^2\right)}\right) \right] - \lambda \sqrt{1-\frac{2 m}{\sqrt{\kappa +r_b^2}}-\frac{1}{3} \Lambda  \left(\kappa +r_b^2\right)} (r_b^2 + \kappa)  \right\}\nonumber\\
    &&\times(1-\tilde{\Delta}).\nonumber\\
\end{eqnarray}
Using the rescaled form of $\beta_0$ previously given in Eq. \eqref{eq:beta0match} as the form above is expanded, we obtain the following results
\begin{eqnarray}
    I_{ct} &=& \frac{\beta_0}{2\kappa} \bigg[-\frac{r^{10} \left(l^5+12 \kappa  l^3-72 \kappa ^2 l\right)}{1728 \kappa ^3 \left(\kappa +r^2\right)^{9/2}} -\frac{r^8 \left(-72 \kappa ^2+l^4+12 \kappa  l^2\right)}{144 \kappa ^2 l \left(\kappa +r^2\right)^{7/2}} + \frac{r^6 \left(-72 \kappa ^2+l^4+12 \kappa  l^2\right)}{24 \kappa  l^3 \left(\kappa +r^2\right)^{5/2}} \nonumber \\
    &&  + \cdots \bigg](1-\tilde{\Delta}).
\end{eqnarray}

The latest approach that we attempt to conduct does not seem to normalize the bulk action, instead, more non-linear divergent terms appear in the counterterm action evaluation. This leads us to conclude that the approach is not effective, and should not be considered as the procedure taken for the counterterm computation.

\end{document}